\documentclass[11pt]{article}

\usepackage{graphicx}
\usepackage{dcolumn}
\usepackage{bm}
\usepackage{epic,eepic}
\usepackage{graphicx}
\usepackage{psfrag}

\usepackage{citesort}
\usepackage{here}

\title{
Exact Solution of a Jamming Transition: Closed Equations for a Bootstrap Percolation Problem}

\author
{Paolo De Gregorio\footnotemark[1], Aonghus Lawlor, Phil
Bradley, Kenneth A. Dawson\\
\\
\normalsize{Irish Centre for Colloid Science and Biomaterials,}\\
\normalsize{Department of Chemistry, University College Dublin,}\\
\normalsize{Belfield, Dublin 4, Ireland}\\
\\
}

\date{}

\begin{document}

\setlength{\textwidth}{26pc}
\baselineskip24pt

\maketitle

\begin{center}
\section*{Abstract}
\end{center}
Jamming, or dynamical arrest, is a transition at which many particles stop moving in a collective manner. In nature it is brought about by, for example, increasing the packing density, changing the interactions between particles, or otherwise restricting the local motion of the elements of the system. The onset of collectivity occurs because, when one particle is blocked, it may lead to the blocking of a neighbor. That particle may then block one of its neighbors, these effects propagating across some typical domain of size named the `dynamical correlation length'. When this length diverges, the system becomes immobile. Even where it is finite but large the dynamics is dramatically slowed. Such phenomena lead to glasses, gels, and other very long-lived non-equilibrium solids. The bootstrap percolation models are the simplest examples describing these spatio-temporal correlations. We have been able to solve one such model in two dimensions exactly, exhibiting the precise evolution of the jamming correlations on approach to arrest. We believe that the nature of these correlations, as well as the method we devise to solve the problem, are quite general. Both should be of considerable help in further developing this field.
\footnotetext[1]{To whom correspondence should be addressed. E-mail: paolo@fiachra.ucd..ie.}

\newpage

\section*{}

There exists within nature a whole class of systems that exhibit a
geometrical `percolation' transition at which they become spanned by a
single infinite cluster extending across the whole
system \cite{kirkpatrick1971,stanley1977,stauffer1992}. Such
transitions may be observed, for example, by randomly occupying
lattice sites at some prescribed density. Spatio-temporal particle
correlations implied by simple dynamical models may also be studied
using percolation ideas. Indeed, since its introduction
\cite{pollak1975,kogut1981}, the potential of the bootstrap
percolation problem \cite{jackle1988,jackle1994} to analyze the dynamics of a
system of highly coupled and locally interacting units has been
recognized. Latterly the range of applications has continued to
grow \cite{kirkpatrick2002,ritort_review,toninelli_jsp2004,pan2004,liu2004}.

This problem is of particular interest because of a growing focus on, and
appreciation of, the unifying role of dynamical
arrest \cite{cocis2000,wolynes1989,mezard2000,stillinger2001,gotze1991},
or `jamming' \cite{liu1998} in the formation of complex condensed
states of matter. Despite many advances, there is as yet no complete
and fundamental conceptual framework to describe the phenomena. In
comparable situations it has been an important lesson of critical
phenomena \cite{onsager,widom1974} that an exact solution, even of a
two dimensional model system, can be of great assistance in broader
efforts to understand the issues. Thus, an exact closed solution of
one bootstrap problem (with all of the implications of strong
packing-induced coupling and divergent correlated domains) would
represent, even without direct access to transport coefficients, a solution of a non-trivial (and non-mean field) `jamming' or
arrest scenario. We will present such a solution in this paper.

That such a treatment is possible must be considered surprising, for
there have been no prior indications of such simplification.

The connection of the bootstrap percolation problem to jamming
phenomena is clear. Thus, particles, processors, or members of a
competing population are considered to become `inactive' (or blocked)
depending on the state of their immediate environment. The relationship of the
bootstrap process to dynamics is as follows. Beginning with a lattice occupied according to the steady state (usually random) distribution of particles of the underlying dynamical model, units not blocked (i.e. movable) are iteratively removed. Those remaining are blocked irrespective of the
arrangements of the movable ones in the configuration. Onset of
extended mutual blocking (in which one unit is prevented from `moving'
by its neighbors and also contributes to the blocking\footnotemark[2] of some of those
neighbors) manifests itself by a sharp change in the occupancy of the
lattice after the bootstrap process. Percolation occurs when, after the iterative procedure finishes, the blocked clusters extend across the whole system. Even on approach to such a transition, single-particle movement requires a long sequential and contingent string of changes in surrounding units. In the underlying dynamics this long string implies a growing correlation length- and accompanying time-scale, both determined by
mutual blocking effects. This (`dynamical') length scale is named the
bootstrap length. Apart from those errors introduced by removal
(rather than movement, and subsequent averaging) of particles, this is
the characteristic blocking or `jamming' length over which motions can
occur only by rare and co-operative motions.  Beyond it, motion is
diffusive.

\section*{Methods}

Bootstrap percolation transitions are believed to fall into two broad
`universality' classes \cite{adler1991}, depending on the dimensionality $d$ and on the local rule `$c$' \footnotemark[2]\footnotetext[2]{In `conventional' bootstrap percolation, particles are removed if they are surrounded by $c$ or fewer particles. In `modified' bootstrap percolation, an additional condition requires the neighboring vacant sites to be second-neighbor to each other.}, and we might expect this to be reflected in the classes of jamming. The first type is a continuous transition in which the blocked domain size diverges as a power law in system size and density. The second type, when $d\leq c < 2d-1$, displays discontinuous transitions with different essential singularities. Specifically, the case $c=d$\footnotemark[2]\cite{kogut1981,lawlor2002prl,toninelli2004} is believed to be more relevant to particle jamming, and there the bootstrap length $\xi$ diverges as $\xi = \exp ^{o(d-1)} (A/(1-\rho))$\cite{cerf2002} (here
$\exp^{o(d-1)}$ is the exponential function iterated $d-1$ times). For
the two-dimensional square lattice ($c=2$)\footnotemark[2], theoretical calculations
based on a set of bounds \cite{holroyd2003} have resulted in an elegant
outcome; essentially exact asymptotic results, $\lim_{ \stackrel{\rho
\to 1}{\xi \to \infty} } 2 (1-\rho) \log \xi=A$, where $A=\pi^{2}/9$
and $\pi^{2}/3$ for conventional and modified
bootstrap\footnotemark[2] respectively. It transpires that at high
packing, this length is also the typical distance between increasingly
rare voids (later named connected holes) in the configuration
\cite{paoloprl2004}, providing a key connection between the
spatio-temporal correlations, and the packing of particle
configurations that determine them.

A striking aspect of this arena of research is that all attempts
\cite{kurtsiefer2003,adler2003} (including some of the most extensive
calculations to have been applied in statistical mechanics) to obtain
agreement between simulations and known exact asymptotic results have
so far failed. Indeed, the bootstrap problem has generated topical
interest \cite{gray-review} because of the notorious (apparent) lack
of agreement between simulation \cite{kurtsiefer2003,adler2003} and
theoretical \cite{aizenman1988,vanenter1990,schonmann1992,holroyd2003,cerf2002}
developments.

In our study we introduce holes or `dynamically available volume'
\cite{lawlor2002prl,faraday}, these being empty sites on the
lattice into which at least one neighboring particle can move. A hole
is termed `connected' or disconnected respectively if the lattice can
(or cannot) be vacated by sequentially removing particles using it as
an origin in the bootstrap process. In the underlying dynamics,
connected holes typically initiate a sequence of moves that will
mobilize every particle in the system. They are therefore the
mediators of transport in the system, and the order parameter of
arrest. For a system to bootstrap (and therefore not be jammed) it
must contain at least one connected hole. We identify $\nu$ as being the connected hole density (average frequency of such occurrences in the lattice). In the high density regime, the bootstrap length may therefore be calculated from $\xi=1/{\nu}^{1/d}$ \cite{paoloprl2004}, the particles within this
distance being movable only via the restricted motions made available
by a single connected hole.

Disconnected holes in the modified bootstrap problem (for $d=2$,
$c=2$-modified\footnotemark[2]) are found to be enclosed by a
single continuous rectangular boundary or `cage' of particles, none of
which can be moved by any particle rearrangements inside the boundary.
They cannot be formed for edge lengths much larger than $l_c$ (the
cut-off size, $l_{c} \sim 1/(1-\rho)$) for then there is typically an
intervening vacancy. We can therefore sample particle removal paths
around holes directly in simulations \cite{paoloprl2004}. If particle
domains much larger than $l_c$ can be removed, the hole may be
identified as connected, rendering the simulation very
efficient. Thus, the ($\circ$) points in the
(Fig. 1) are hole densities one would obtain for
systems sizes up to $L=220,000$. They constitute the largest
simulations to have been carried out on a two-dimensional bootstrap
problem.

\begin{figure}[ht!]
\includegraphics[angle=90,width=\columnwidth,height=!]{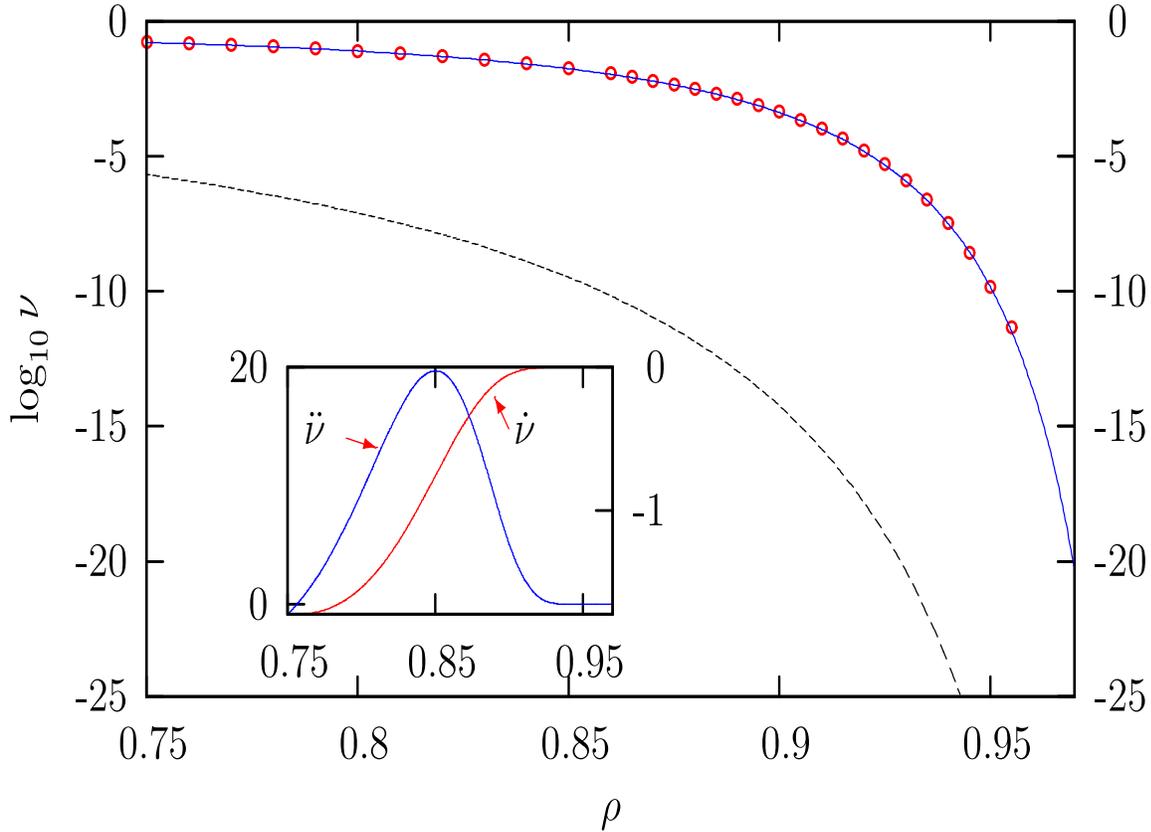}
\caption{Connected hole density in the Modified Bootstrap Model. Main plot: The ($\circ$) points represent the total hole density from simulations (\cite{paoloprl2004}), the last point being equivalent to a system of size 220,000, the current limit of computer simulation. The solid line represents the result of our new exact theory. These are compared with the asymptotic result $\exp(-\pi^{2}/3(1-\rho))$ (dashed line) \cite{holroyd2003}. We recall that results for the total connected hole density $\nu$ coincide with $1/\xi^2$ at high density\cite{paoloprl2004}, where $\xi$ is the bootstrap percolation correlation length. Inset: $\partial\nu/\partial\rho$ (Red) and $\partial^2\nu/\partial\rho^2$ (Blue) against density in the same range. $\partial\nu/\partial\rho$ increases sharply from $-1.73$ to approximately zero with an inflexion signalled by the peak in $\partial^2\nu/\partial\rho^2$ at $\rho=0.85$.}
\end{figure}

The bootstrap problem is an example of strong (packing-induced)
coupling, so theoretical developments, including simple extensions of
previous approaches \cite{holroyd2003} involving asymptotic bounds,
invariably fail. Instead we define new intermediate states that imply
bootstrap-type (ie spatio-temporal) correlations. Transitions between
them may then be composed into sequences or particle `removal paths'
whose steps are de-correlated, and transition coefficients
($c^{(i,j)}_{\bf k}(\rho)$) from state to state are then easily
calculated from the random ensemble. These intermediate states are
rectangular regions (of size $k$ by $k+s$, see
Fig. 2). They are, irrespective of their orientation
in space, completely empty, or empty save for one, two (adjacent) or
three edges occupied entirely by particles. The probabilities that
such rectangles are visited during the emptying process are
$P^{(i)}_{k,k+s}(\rho)$, $i=1,4$ (the states $P^{(2)}_{k+s,k}(\rho)$ and $P^{(4)}_{k+1,k}(\rho)$ are also defined, these bearing the same meaning as $P^{(2)}_{k,k+s}(\rho)$ and $P^{(4)}_{k,k+1}(\rho)$, but for the orientation of the occupied lines with respect to the direction of the elongation). In defining the transitions between
them, only certain directions (illustrated by arrows in
Fig. 2) of the removal paths are permitted\footnotemark[3]\footnotetext[3]{These movements tend to restore the symmetric empty square, except for the state represented by $P^{(2)}_{k,k+s}(\rho)$.}. For example, if the path leads to a state `1'
(probability $P^{(1)}_{k,k+s}(\rho)$) the next attempt to grow the
vacated rectangle will be via either of the two `long' directions,
this requiring removal of particles along one or other of those
directions. These directions are chosen randomly, with equal
probability. The next intermediate state will be a $k+1$ by $k+s$
rectangle, either in state `1' or in state `2', since these represent
the only outcome of that transition. The transition (`jump') rates for
these two choices are $(1-\rho^{k+s})$ and $\rho^{k+s}$, and they
define $c^{(1,1)}_{k+1,k+s}(\rho)$ and $c^{(2,1)}_{k+1,k+s}(\rho)$
respectively.

\begin{figure}[ht!]
\includegraphics[width=\columnwidth,height=!]{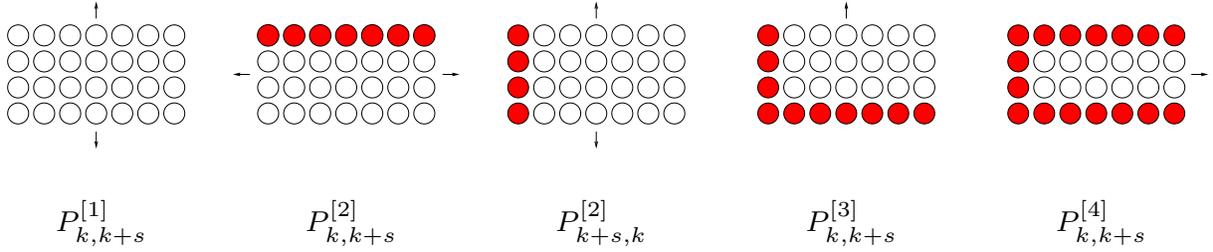}
\caption{Intermediate states. Intermediate local configurations surrounding a candidate hole required in evaluation of the sum over `particle removal paths' that are used to empty the lattice ($k=4$ and $s=3$ in the example). The white circles are sites that have been emptied in previous steps, and the red disks are temporarily blocked particles along the perimeter edges. For each configuration, the arrows indicate allowed transitions (to be selected at random) involving the growth of one boundary line by one step. This process restores that local configuration to another intermediate state ($P^{(i)}_{k+1,k+s}$ or $P^{(i)}_{k,k+s+1}$) with the indices $k,s$ suitably revised. The process terminates (as a disconnected hole) when the local configuration makes a transition from $P^{(4)}_{k,k+s}$ outside of the class $P^{(i)}_{k,k+s+1}$.}
\end{figure}

We reiterate these choices. In essence a highly correlated bootstrap
process becomes uncorrelated within the subset of intermediate states,
leading to an exact method to identify connected holes by random
sampling of a constrained set of states. For connected holes we remain
within the closed set of rectangular states, with the indices suitably
modified after each step. Disconnected holes, being bounded by a
rectangular contour of particles, are identified by a transition
outside of this set of states. No transition between states involves
more than one attempt to remove a group of particles, and the
transition rates may be calculated from the random ensemble. The
equations of the bootstrap probabilities, if one recalls that in the
bootstrap problem movable and removable particles are analogous,
represent a map of dynamics at one scale onto the next largest scale,
with the packing correlations automatically accommodated.

These features constitute the key elements of the methodological
advances described in this paper, and similar treatments may be
attempted for other such problems in future \cite{paoloprl2004}.

The exact equations representing the process may be written in closed
form,

\begin{eqnarray}
P ^{(1)} _{k,k+s}  &=& (1-\rho ^{k+s})P ^{(1)} _{k-1,k+s} + (1-\delta _{s, 0})(1-\rho) P ^{(2)} _{k,k+s-1} \label{exact} \\
&&+ (1-\rho)P ^{(2)} _{k+s,k-1}  + \delta _{s,1}(1-\rho ^k) P ^{(1)} _{k,k}  \nonumber \\
P ^{(2)} _{k,k+s}  &=& \rho ^{k+s}P ^{(1)}_{k-1,k+s} + \rho(1-\rho ^{k-1})  P ^{(2)} _{k,k+s-1} + (1-\rho) P ^{(3)} _{k-1,k+s} + \delta
_{s,0} (1-\rho) ^2 P ^{(4)}
_{k,k-1}  \nonumber \\
P ^{(2)} _{k+s,k}  &=& \delta _{s,1}
{\big{[}} \rho ^k P ^{(1)} _{k,k}  + (1-\rho) P ^{(3)} _{k,k} {\big{]}} \nonumber \\
&&+ (1-\delta_{s,0}) {\big{[}}\rho (1-\rho ^{k+s-1}) P ^{(2)} _{k+s,k-1}
+ (1-\rho) ^2 P ^{(4)} _{k,k+s-1} {\big{]}} \nonumber \\
P ^{(3)} _{k,k+s} &=& (1-\delta _{s, 0}) \rho^k P ^{(2)} _{k,k+s-1}+ \rho
^{k+s} P ^{(2)} _{k+s,k-1} + \rho (1-\rho ^{k+s-1}) P^{(3)} _{k-1,k+s} \nonumber \\
&&+ 2 \rho(1-\rho) P ^{(4)} _{k,k+s-1} +
\delta_{s,1}\rho(1-\rho^{k-1})P ^{(3)} _{k,k} \nonumber \\
P^{(4)} _{k,k+s} &=& \rho^{k+s} P ^{(3)} _{k-1,k+s} + \rho^2(1-\rho^{k-2}) P^{(4)} _{k,k+s-1} \nonumber \\
P ^{(4)}_{k+s,k}&=&\delta_{s,1}\rho^k P ^{(3)} _{k,k}. \nonumber
\end{eqnarray}
Here $s\geq 0$ and the equations are solved numerically subject to the
initial conditions, $P_{1,k}^{(i)}=(1-\rho)\delta_{1i}
\delta_{1k}$. These equations converge numerically for $k \gg l_{c}$
and $0 < s/k < 1$.  The connected hole density, the probability to
bootstrap a square of indefinite size, is then given by the
expression,
\begin{eqnarray}
\nu(\rho) = \lim_{k\rightarrow \infty} P ^{(1)} _{k,k}(\rho).
\end{eqnarray}

Because of the cut-off in cage size, these equations converge very
rapidly and there are no practical limitations on the density range
that can be studied by computer, as already discussed in
\cite{paoloprl2004}. Based on exact asymptotic analysis, the quantity
$-2(1-\rho)\log\xi+\pi^2/3=f(\rho)$ would be predicted to be positive
and vanishing in the asymptotic limit ($\rho\rightarrow 1$). Now we have
numerically calculated the solution for the modified model up to the
largest sizes allowed by the precision of the computer, and indeed
this quantity does decrease monotonically when $\rho$ approaches
$1$. However, at $\rho \sim 0.9965$, $f(\rho)$ is still of order
unity, rather too large to be neglected. Suffice it to say that the
true exact bootstrap length for this density is $\xi\sim
10^{150}$. Thus in reality, for large but finite systems one sees
$\xi=\exp[-f(\rho)/2(1-\rho)]\exp[\pi^2/6(1-\rho)]$, $f(\rho)$ being a
decreasing but `always' finite function of $(1-\rho)$. Equations (\ref{exact}) may also be solved
asymptotically. We illustrate in Fig. 3 the density
of cages (dimension $k$ by $k+s$) at particle density $\rho=0.95$ and
the probability for vacated square-like regions $P ^{(1)}
_{k,k}(\rho)$. There we see that cages become rare somewhat beyond the
critical size $l_c$\footnotemark[4]\footnotetext[4]{The threshold value $l_c$ in (Fig. 3) can be understood as that size at which, with equal probability, the droplet can either grow or find a cage in the next step.}. The reduction of cages, capable
of preventing bootstrapping, leads to the dominance of $P ^{(1)}
_{k,k}(\rho)$ and $P ^{(1)} _{k,k+1}(\rho)$ beyond the critical core
size.

\begin{center}
\begin{figure}[ht!]
\includegraphics[angle=90,width=0.9\columnwidth,height=!]{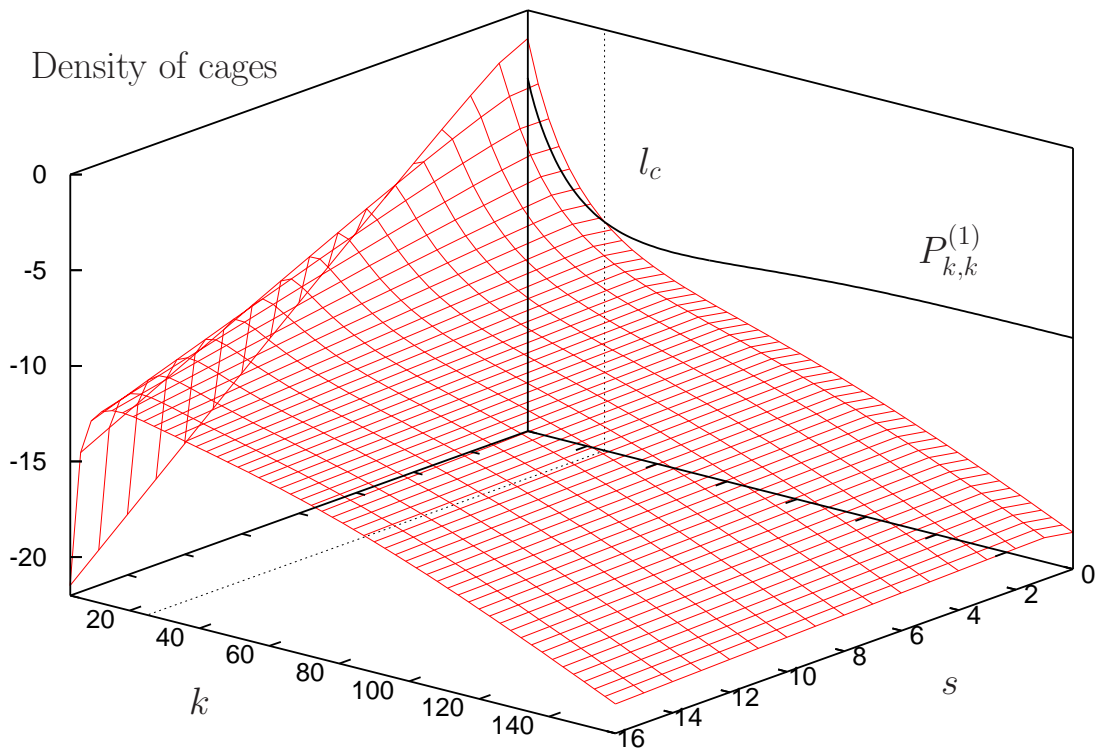}
\caption{Cages. Particle density $\rho=0.95$. Illustrative calculation in the `metastable' or droplet state. The red surface represents the probability (in $\log_{10}$ scale) that a rectangular cage surrounding a vacant site is first found at size $k$ by $k+s$. The probability density of cages is plotted against $k$ (core size) and $s$ (asymmetry-degree), and decreases monotonically in $k$. High degrees of asymmetry are very rare. The probability of finding an empty square of size $k$ is also plotted for a comparison. The intersection with the corresponding cage size distribution is at $l_c=25$
, indicating that cages become statistically irrelevant (i.e. extremely rare) already at sizes much smaller than $\xi \sim 10^5 $.}

\end{figure}
\end{center}

\section*{Results}

The results for connected hole density, the full curve in
Fig. 1, or equivalently the bootstrap length,
settle a longstanding mystery, the theory now being in precise
agreement with simulation where the latter is applicable. They permit
us to explore the nature of the problem all the way from lower density
to the asymptotic form, as solutions to a single closed set of
equations. It is now clear that there has never been any inconsistency
between the simulations and asymptotic results. They were both
correct, but direct comparison is not appropriate for any range of
density appropriate to computer simulation. The numerical results (and
associated asymptotic analysis) from these same equations are
sufficient to show that the previously reported exponential asymptotic
form \cite{holroyd2003} occurs for such high density (and large
length-scales) that systems of physical interest will never reach this
regime. As noted in the previous parapraph, even at system sizes of $\xi\sim 10^{150}$ there remains a
singular model-dependent pre-factor to the leading
exponential. Formulae purporting to
describe the behavior for finite systems (or systems at finite
density) in terms of a simple exponential are effective fits, and
possess no fundamental significance.

The real paradoxes lie in the Physics. As we now show, this Physics
turns out to be much richer than had been supposed until now. In
discussing these issues we have in mind primarily the clarification of
the model system, though the results may relate to recent experimental
observations, and could help to clarify these issues also. We give
only pointers to the new directions.

Recall that the connected hole density in a region characterizes the
ease of long-ranged movement of particles in that region, these being
the empty spaces on the lattice that permit sustained motion. Given
this, the insets to (Fig. 1) are striking. There
we show both the first
(${\dot{\nu}}=\partial\nu/\partial\rho=-\partial\nu/\partial v$ ) and
second derivatives
(${\ddot{\nu}}=\partial^2\nu/\partial\rho^2=\partial^2\nu/\partial
v^2$) of the connected hole density with respect to particle (or
vacancy, $v$) density. The former is the long-wavelength limit of the
response of dynamical processes in the system to small changes in
particle density. We observe a `transition' in the response of the
dynamics to small changes in particle density. For densities lower
than $\rho=0.85$ (the `unstable' regime\footnotemark[5]\footnotetext[5]{The low density near-arrest regime has a hole density that is well fitted to power laws $(\rho_c-\rho)^\mu$ with $\rho_c$ and $\mu$ suitably defined in the range of interest, whilst the high density regime is well-fitted to the exponential law $\exp[f(\rho)/(1-\rho)]\exp[-\pi^2/3(1-\rho)]$.}) a small
local decrease of density leads to many new connected holes, and
thereby opens up many new transport pathways. For particle densities
higher than this (the `metastable' regime) the response of dynamical
properties to a density fluctuation is vanishingly small.

Albeit in this quite new context of near-arrested systems, these ideas
make contact with early prescient remarks \cite{aizenman1988} in which
it was suggested that bootstrap models might offer an explanation for
the decay of the metastable state and (by extension) the `spinodal'
transition to unstable state decay. Indeed, the two regimes of
connected hole density reflect the relaxation of particle
configurations by two characteristic means. At low density,
configurations are intrinsically unstable to small spatial variations
of density, these leading to new dynamical pathways reminiscent of
`spinodal waves' that enable efficient relaxation of quite dense
particle configurations. Beyond the peak at $\rho=0.85$ the state
develops metastability, this being reflected also in the
characteristic exponentially small connected hole density there, and
the decay of such configurations becomes much more difficult. In (Fig.
4) we illustrate these ideas in a more visual
manner. Thus, the left picture (Fig. 4)
represents a configuration at density $\rho=0.79$. There we see the
development of `spinodal-like' patterns of empty space as particles
are removed from the unstable state. The right picture
(Fig. 4), at $\rho=0.915$ exhibits the classical
appearance of decay of the metastable state, including growing empty
droplets, with smaller sub-critical droplets failing to grow any
further. These analogies will prove deeply helpful in future
formulations of quantitative laws for dynamical arrest also
\cite{lawlor2004u}.

\begin{figure}[ht!]
\includegraphics[width=\columnwidth,height=!]{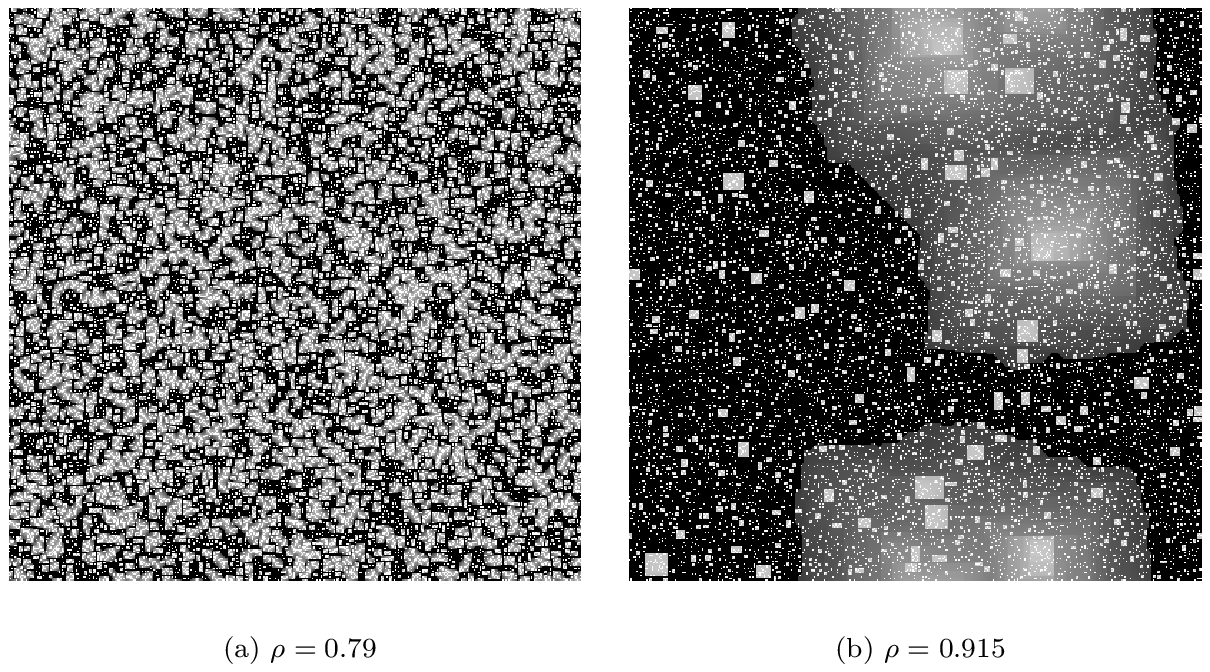}
\caption{The two regimes. Configurations ($L=400$) after partial removal of particles via parallel updating. The two examples refer to the `unstable' regime (a) and to the `metastable' regime (b), lying respectively to the left and right of the peak in the inset of Fig. 1. The process is stopped after removal of $50\%$ of the particles (black) (the grayscale representing the time during the process when the particles were removed).}
\end{figure}

The peak of the second derivative in (Fig. 1)
defines a transition between two near-jammed regimes, the first
`unstable' regime where spatial density heterogeneities produce large
heterogeneities also in the dynamics, the second `metastable' regime
where the response is much weaker. This observation may well be
related to recent ideas derived from
simulations \cite{franz2000a,glotzer2000a} and
experiments \cite{weeks2000,vanblaaderen2000,cipelletti2003} where
dynamically heterogeneous processes have been viewed as intrinsic to
the arrest process. The `size' of these domains in our bootstrap
calculation ($1/\sqrt{\nu}$) in the `unstable' regime reaches a
maximum value of $l_d=8$ at $\rho =0.85$. Thereafter the dynamically
heterogeneous processes change their nature to `droplets' of movement
and the system continues to slow further, as the number of connected
holes decreases. Thus, the dynamical (here bootstrap) correlation
length continues to diverge, but no longer has the same quality of
`dynamical heterogeneity'.

That there exists two such regimes, with a clearly defined transition
between them has not been previously clarified in the scientific
literature. There is inferential evidence from simulations and
experiment that dynamical heterogeneities near
`arrest' \cite{glotzer2000a} might change as the system further
progresses towards complete arrest. However, the broader implication
of this discussion is that dynamical heterogeneities, and the
`metastable' state, could be formulated into a theoretical
approach via these response functions in available volume. It is
encouraging to note that, with the advent of new microscopy
tools \cite{vanblaaderen2000,weeks2000}, there has emerged the
possibility to directly measure these new more appropriate order
parameters and response functions using direct optical imaging.

We do not here deal with the issue in detail, but note that,
associated with the two regimes of the response function, the
transport coefficients governing these two regimes should be quite
different. This observation might also point to a resolution of
another long-standing controversy about the nature of the dynamical
laws (power law or `exponential') near arrest\footnotemark[5].

\section*{Discussion}

In summary, a new approach to treating strong jamming correlations
using random paths between appropriately selected states means that
the two-dimensional bootstrap model can now be solved exactly. This advance
allows us to explore phenomena far beyond the reach of simulations
using modern computers, and provides us with the first complete
picture of the onset of arrest\footnotemark[6]\footnotetext[6]{It is possible to show that more complicated jamming scenarios may also be exactly solved in two dimensions providing the defining constraints on particle movement remain short-ranged on the lattice.}. The conclusions
of the study are intriguing and resolve long-standing apparent
conflicts between theoretical and simulation treatment of these
problems.

True arrest occurs only as a limit of a `metastable state' when the
system is dense-packed on the lattice. Thus, even the smallest amounts
of available volume means that motion is still possible, albeit at
enormously long time-scales. The finite size dependence of this regime
is barely appreciable, but nevertheless persists up to the fully
packed limit and means that neither simulations, nor nature, can ever
access length scales where simple asymptotic results in terms of
density are valid\footnotemark[7]\footnotetext[7]{This issue deserves some comment. There is as yet not settled opinion on where, for continuum systems, true dynamical arrest occurs. The continuum analogue of the lattice arrest transitions would likely be random close packing for hard spheres, or the Kautzmann condition for systems with finite energy repulsions.}. However, the results
written in terms of connected hole density, or dynamically available
volume, are simple. This has deep implications for the way in which
experiments (indeed the whole study of dynamical arrest) might be
framed in future \cite{lawlor2002prl,faraday}.

Finally, and important to the general arena of dynamical arrest, we
have shown how the conceptual infrastructure of order parameters,
dynamically available volumes, spatio-temporal correlation lengths,
and equations to represent them, may be framed into a closed theory,
from which rational approximations emerge. However, it is our belief
that these new principles governing treatment of the bootstrap are
close to those required to frame a comparable dynamical theory, from
which one would obtain the transport coefficients in systems that are
dynamically slowed.

Understanding and ultimately calculating these transport coefficients
in dynamically slowed systems from a fundamental theoretical basis,
and comparing them to results from carefully designed experiments,
constitutes one of the outstanding challenges of those interested in
the field of dynamical arrest.

\section*{Acknowledgements}
We acknowledge fruitful interactions with A. van Enter, E. Marinari, M. Mezard, G. Parisi, A. Robledo, M. Sellitto, D. Stauffer, P. Tartaglia.


\begin{thebibliography}{10}

\bibitem{kirkpatrick1971}
Shante, V. K.~S \& Kirkpatrick, S.
\newblock (1971) {\em Adv. in Phys.} {\bf 30}, 325--357.

\bibitem{stanley1977}
Reynolds, P.~J, Klein, W,  \& Stanley, H.~E.
\newblock (1977) {\em J. Phys. C} {\bf 10}, L167--L172.

\bibitem{stauffer1992}
Stauffer, D \& Aharony, A.
\newblock (1992) {\em Introduction to Percolation Theory}.
\newblock (Taylor and Francis).

\bibitem{pollak1975}
Pollak, M \& Reiss, I.
\newblock (1975) {\em Phys. Status Solidi b} {\bf 69}, K15--K18.

\bibitem{kogut1981}
Kogut, P.~M \& Leath, P.~L.
\newblock (1981) {\em J. Phys. C.} {\bf 14}, 3187--3194.

\bibitem{jackle1988}
Ertel, W, Frobr\"ose, K,  \& J\"ackle, J.
\newblock (1988) {\em J. Chem. Phys.} {\bf 88}, 5027--5034.

\bibitem{jackle1994}
J\"ackle, J \& Kr\"onig, A.
\newblock (1994) {\em J. Phys.: Condens. Matter} {\bf 6}, 7633--7653.

\bibitem{kirkpatrick2002}
Kirkpatrick, S, Wilcke, W, Garner, R,  \& Huels, H.
\newblock (2002) {\em Physica A} {\bf 314}, 220--229.

\bibitem{ritort_review}
Ritort, F \& Sollich, P.
\newblock (2003) {\em Adv. in Phys.} {\bf 52}, 219--342.

\bibitem{toninelli_jsp2004}
Toninelli, C \& Biroli, G.
\newblock (2004) {\em J. Stat. Phys.} {\bf 117}, 27--54.

\bibitem{pan2004}
Pan, A.~C, Garrahan, J.~P,  \& Chandler, D.
\newblock (2004) {\em cond-mat/0410525}.

\bibitem{liu2004}
Schwarz, J.~M, Liu, A.~J,  \& Chayes, L.~Q.
\newblock (2004) {\em cond-mat/0410595}.

\bibitem{cocis2000}
Dawson, K.~A.
\newblock (2000) {\em Curr. Opin. in Coll. \and Int. Sci.} {\bf 7}, 218--227.

\bibitem{wolynes1989}
Kirkpatrick, T.~R, Thirumalai, D,  \& Wolynes, P.~G.
\newblock (1989) {\em Phys. Rev. A} {\bf 40}, 1045--1054.

\bibitem{mezard2000}
Mezard, M \& Parisi, G.
\newblock (2000) {\em J. Phys. Condens. Matter} {\bf 12}, 6655--6673.

\bibitem{stillinger2001}
Debenedetti, P.~G \& Stillinger, F.~H.
\newblock (2001) {\em Nature} {\bf 410}, 259--267.

\bibitem{gotze1991}
G\"otze, W.
\newblock (1991) in {\em Freezing and Glass Transition}, eds.{} Hansen, J.~P,
  Levesque, D,  \& Zinn-Justin, J.
\newblock (Amsterdam: North Holland), pp. 287--343.

\bibitem{liu1998}
Liu, A.~J \& Nagel, S.~R.
\newblock (1998) {\em Nature} {\bf 396}, 21--22.

\bibitem{onsager}
Onsager, L.
\newblock (1944) {\em Phys. Rev.} {\bf 65}, 117--149.

\bibitem{widom1974}
Widom, B.
\newblock (1974) {\em Physica} {\bf 73}, 107--118.

\bibitem{adler1991}
Adler, J.
\newblock (1991) {\em Physica A.} {\bf 171}, 453--470.

\bibitem{lawlor2002prl}
Lawlor, A, Reagan, D, McCullagh, G.~D, {De Gregorio}, P, Tartaglia, P,  \&
  Dawson, K.~A.
\newblock (2002) {\em Phys. Rev. Lett.} {\bf 89}, 245503.

\bibitem{toninelli2004}
Toninelli, C, Biroli, G,  \& Fisher, D.
\newblock (2004) {\em Phys. Rev. Lett.} {\bf 92}, 185504.

\bibitem{cerf2002}
Cerf, R \& Manzo, F.
\newblock (2002/9) {\em Stochastic Processes and their Applications} {\bf 101},
  69--82.

\bibitem{holroyd2003}
Holroyd, A.
\newblock (2003) {\em Probability Theory and Related Fields} {\bf 125}, 195--226.

\bibitem{paoloprl2004}
{De Gregorio}, P, Lawlor, A, Bradley, P,  \& Dawson, K.~A.
\newblock (2004) {\em Phys. Rev. Lett..} {\bf 93}, 025501.

\bibitem{kurtsiefer2003}
Kurtsiefer, D.
\newblock (2003) {\em Int. J. Mod. Phys. C.} {\bf 14}, 529--536.

\bibitem{adler2003}
Adler, J \& Lev, U.
\newblock (2003) {\em Brazilian Journal of Physics} {\bf 33}, 641--644.

\bibitem{gray-review}
Gray, L.
\newblock (2003) {\em Notices of the AMS} {\bf 50}, 200--211.

\bibitem{aizenman1988}
Aizenman, M \& Lebowitz, J.~L.
\newblock (1988) {\em J. Phys. A: Math. Gen.} {\bf 21}, 3801--3813.

\bibitem{vanenter1990}
{van Enter}, A. C.~D, Adler, J,  \& Duarte, J. A. M.~S.
\newblock (1990) {\em J. Stat. Phys.} {\bf 60}, 323--332.

\bibitem{schonmann1992}
Schonmann, R.
\newblock (1992) {\em Ann. Probab.} {\bf 20}, 174--193.

\bibitem{faraday}
Dawson, K.~A, Lawlor, A, {De Gregorio}, P, McCullagh, G.~D, Zaccarelli, E,
  Foffi, G,  \& Tartaglia, P.
\newblock (2003) {\em Faraday Discussions} {\bf 123}, 13--26.

\bibitem{lawlor2004u}
Lawlor, A, {De Gregorio}, P, Bradley, P, Sellitto, M,  \& Dawson, K.~A.
\newblock (2005) {\em cond-mat/0503089}.

\bibitem{franz2000a}
Franz, S \& Parisi, G.
\newblock (2000) {\em J. Phys.: Condens. Matter} {\bf 12}, 6335--6342.

\bibitem{glotzer2000a}
Glotzer, S.~C, Novikov, V.~N,  \& Schroeder, T.~B.
\newblock (2000) {\em J. Chem. Phys.} {\bf 112}, 509--512.

\bibitem{weeks2000}
Weeks, E.~R, Crocker, J.~C, Levitt, A.~C, Schofield, A,  \& Weitz, D.~A.
\newblock (2000) {\em Science} {\bf 287}, 627--631.

\bibitem{vanblaaderen2000}
Kegel, W.~K \& van Blaaderen, A.
\newblock (2000) {\em Science} {\bf 287}, 290--293.

\bibitem{cipelletti2003}
Cipelletti, L, Bissig, H, Trappe, V, Ballesta, P,  \& Mazoyer, S.
\newblock (2003) {\em J. Phys.: Cond. Matt.} {\bf 15}, S57--S64.

\end{thebibliography}

\end{document}